\documentclass[journal]{IEEEtran}

\usepackage{graphicx}
\graphicspath{{./}{./images/}{./images_all/}{./images_all/analog/}}
\DeclareGraphicsExtensions{.eps}

\usepackage[cmex10]{amsmath}

\usepackage{subfigure}

\begin{document}
\title{Ultralow Energy Analog Straintronics Using Multiferroic Composites}

\author{\IEEEauthorblockN{Kuntal Roy}
\thanks{Kuntal Roy is with the School of Electrical and Computer Engineering, Purdue University, West Lafayette, IN 47907, USA (email: royk@purdue.edu).}
\thanks{This work was supported by FAME, one of six centers of STARnet, a Semiconductor Research Corporation program sponsored by MARCO and DARPA.}}

\markboth{Ultra-low-energy analog straintronics using multiferroic composites}%
{Roy, K.}

\maketitle

\begin{abstract}
Electric field-induced magnetization switching in multiferroics holds profound promise for ultra-low-energy computing in beyond Moore's law era. Bistable nanomagnets in the multiferroics are usually deemed to be suitable for storing a binary bit of information and switching between the two stable states allows us to process digital information. However, it requires to process continuous analog signals too for seamless integration of nanomagnetic devices in our future information processing systems. Here, we show that it is possible to harness the analog nature in the magnetostrictive nanomagnets, contrary to writing a digital bit of information. By solving stochastic Landau-Lifshitz-Gilbert equation of magnetization dynamics at room-temperature, we demonstrate such possibility and show that there exists a transistor-like high-gain region in the input-output characteristics of the magnetostrictive nanomagnets in strain-mediated multiferroic composites. This can be the basis of ultra-low-energy analog and mixed signal precessing in our future information processing systems and it eliminates the requirement of using charge-based transistors.
\end{abstract}


\section{Introduction}
\label{sec:introduction}

\IEEEPARstart{T}{he} invention and development of traditional charge-based transistor electronics has been a story of great success~\cite{RefWorks:553}, however, fundamental limits hindrance the further progress~\cite{RefWorks:211}. In quest of energy-efficient computing, electron's spin-based counterpart, so-called spintronics, has been widely studied in the context of nanomagnets~\cite{RefWorks:557,RefWorks:148,RefWorks:149,roy13_spin,RefWorks:435,RefWorks:328,RefWorks:322,roy14_6,roy14_3,RefWorks:180}. Nanomagnets with two stable states separated by an energy barrier are usually envisaged by the spintronics community for non-volatile binary switching~\cite{roy13_spin}. While switching the magnetization direction between its two stable states allows us to process information digitally (encoded as \emph{0} and \emph{1})~\cite{roy13_spin,roy13,roy14}, it requires to perform \emph{analog} signal processing~\cite{razav01} using the inherently bistable nanomagnets for seamless integration of nanomagnetic devices on a chip. The analog signal processing capability is sometimes fundamentally necessary too, e.g., while processing natural signals that are attenuated in the environment and it needs amplification before further processing in digital domain~\cite{razav01}. Such voltage gain is also necessary for error resiliency in digital integrated circuits~\cite{rabae03}.

\begin{figure}
\centering
\includegraphics{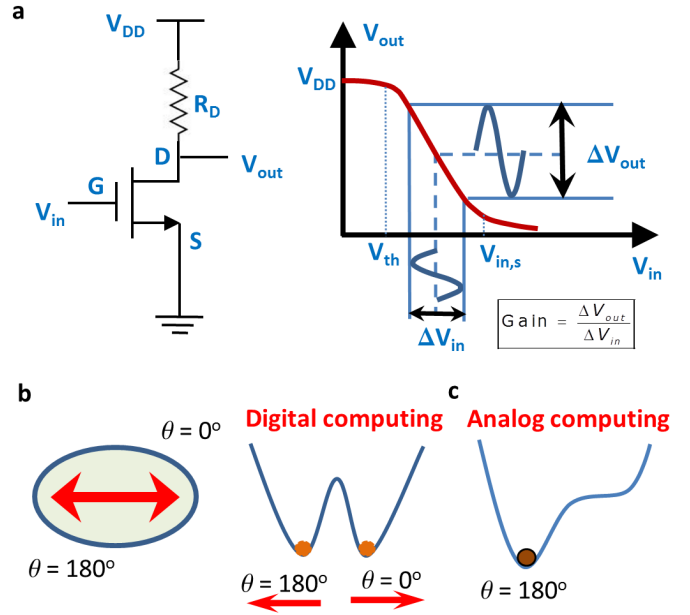}
\caption{\label{fig:roy_fig1} {Analog signal processing using traditional transistors and a proposal of analog computing using shape-anisotropic single-domain nanomagnets.} 
\textbf{(a)} Input voltage ($V_{in}$) versus output voltage ($V_{out}$) characteristics of a traditional $n$-type metal-oxide-semiconductor field-effect-transistors (MOSFETs). When the input voltage $V_{in}$ is applied at the Gate (G) of the transistor, a current flows through the source (S) and drain (D), and the output voltage $V_{out}$ is extracted from the drain. The source is grounded and the drain is connected to the supply voltage $V_{DD}$ through a resistor $R_D$. Around the threshold voltage $V_{th}$, the current starts to flow significantly and at the saturation input voltage $V_{in,s}$, the current flow saturates. In the transition region, an input voltage can be biased and an output voltage can be obtained with a gain of $\Delta V_{out}/\Delta V_{in}$.
\textbf{(b)} A shape-anisotropic single-domain nanomagnet with two $180^\circ$ symmetry equivalent states and such two stable easy axes are shown in the potential landscape of the nanomagnet. Switching between the two stable states allows us to store a non-volatile binary bit of digital information. 
\textbf{(c)} To harness the analog computing capability and facilitate a continuous rotation of magnetization, we propose to make the potential landscape of a nanomagnet monostable to avoid any abrupt switching of magnetization.
}
\end{figure}

\begin{figure*}
\centering
\includegraphics[width=\textwidth]{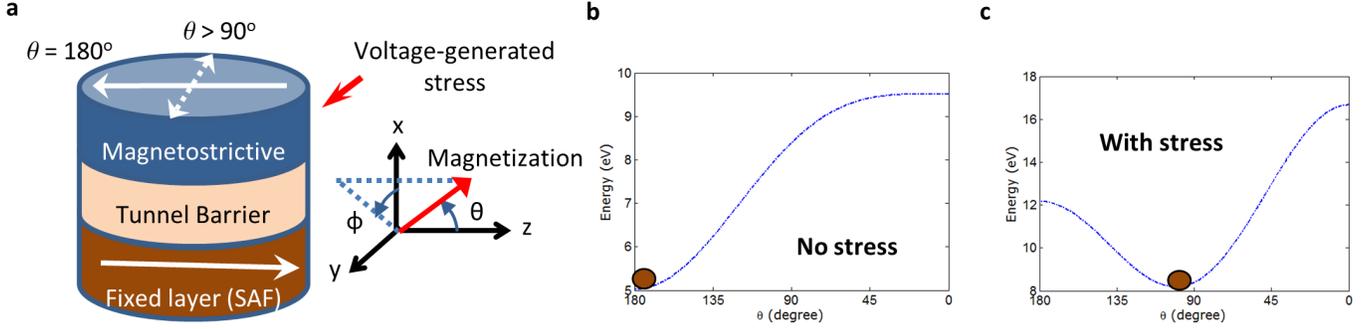}
\caption{\label{fig:potential_profile_magnetostrictive} {A magnetostrictive nanomagnet with an effective field exerted from a fixed layer and its potential landscape without and with application of stress.} 
\textbf{(a)} A single-domain magnetostrictive nanomagnet lying on the $y$-$z$-plane. The fixed layer is exerting an effective field $H_{fixed}$ on the magnetostrictive nanomagnet.
\textbf{(b)} If $H_{fixed}=H_k$, the potential landscape of the magnetostrictive nanomagnet becomes monostable. Here, the minimum energy position is $\theta=180^\circ$. For $H_{fixed} > H_k$, the minimum energy position would be $\theta=180^\circ$ too.
\textbf{(c)} A stress of 30 MPa is applied on the Terfenol-D magnetostrictive nanomagnet to shift its minimum energy position around $\theta=90^\circ$. For increasing value of $H_{fixed}$ over $H_k$, it would require an increasing amount of stress to make the minimum energy position close to $\theta=90^\circ$.
}
\end{figure*}

Figure~\ref{fig:roy_fig1}a depicts the basic notion of processing analog signals using traditional transistors~\cite{razav01}. When the input voltage is higher than the threshold voltage, a significant amount of current starts to flow and when a sufficient input voltage is applied, the current saturates. While the high- and low-state encapsulate a binary bit of \emph{digital} information, the transition region in between is gradual but has a high slope, where we can bias a continuous \emph{analog} input voltage and generate an output voltage with a voltage gain. Figure~\ref{fig:roy_fig1}b shows the potential landscape of a nanomagnet with two anti-parallel stable states storing a binary bit of information $0$ and $1$. For digital computing, we need to switch the magnetization between these two stable states. However, for analog computing, we select one potential well to operate by making the potential landscape monostable since we want to restrict abrupt switching from one stable state to another. This is depicted in the Fig.~\ref{fig:roy_fig1}c with a monostable potential well. 

Recently, it is shown that electric field induced magnetization switching in strain-mediated multiferroic composites, i.e., a piezoelectric layer strain-coupled with a magnetostrictive nanomagnet~\cite{RefWorks:164,RefWorks:165,RefWorks:519}, dissipates a miniscule amount of energy of $\sim$1 attojoule (aJ) in sub-nanosecond switching delay at room-temperature~\cite{roy13_spin,roy11_news,roy11_2,roy13_2,roy11_6,roy14_5}. Also, the Landauer limit of energy dissipation~\cite{RefWorks:148} can be achieved in these magnetostrictive nanomagnets~\cite{roy14_4}. The induced stress anisotropy in magnetostrictive nanomagnets has been experimentally demonstrated too~\cite{RefWorks:559,RefWorks:806,RefWorks:836,RefWorks:609,RefWorks:790,RefWorks:838,RefWorks:868}. With voltage generated stress in a magnetostrictive nanomagnet, the potential landscape of the nanomagnet can be modulated and thereby here we show that it is possible to harness the analog signal processing capability~\cite{roy_aps_2014,roy_mrs_2014,roy_spie_2014}. By solving stochastic Landau-Lifshitz-Gilbert equation~\cite{RefWorks:162,RefWorks:161,RefWorks:186} of magnetization dynamics in the presence of room-temperature thermal fluctuations, we show that a \emph{continuous} rotation of magnetization is possible and a transistor-like high-gain region in the input-output characteristics of these multiferroic composite devices can be achieved~\cite{roy_aps_2014,roy_mrs_2014,roy_spie_2014}. A voltage gain of 50 can be achieved while dissipating only 0.1 aJ/cycle at 1 GHz frequency. 

Being able to perform analog signal processing too using these multiferroic devices will facilitate seamless integration of nanomagnetic devices in the same manufacturing process on chips and will eliminate transistors as supporting devices. Therefore, this is key to the success of building energy-efficient technology based on these multiferroic devices.

\section{Model: Solution of stochastic Landau-Lifshitz-Gilbert (LLG) equation of magnetization dynamics in the presence of thermal fluctuations}
\label{sec:stochastic_LLG}

We adopt the standard spherical coordinate system (see Fig.~\ref{fig:potential_profile_magnetostrictive}a) to solve the stochastic LLG equation of magnetization dynamics. The magnets are lying on the $y$-$z$ plane, and $x$ is the out-of-plane direction. In standard spherical coordinate system, $\theta$ is the polar angle and $\phi$ is the azimuthal angle.

First, we will consider that a magnetostrictive nanomagnet on which an effective magnetic field is exerted from the fixed layer in an magnetic tunnel junction (MTJ) structure (see Fig.~\ref{fig:potential_profile_magnetostrictive}a). This structure is used to explain the concepts behind the \emph{continuous} rotation of magnetization in the magnetostrictive nanomagnet with the application of stress. There is a high gain region in the stress versus the mean orientation of magnetization, where we can bias an AC signal. Therefore, stress is time-dependent in the formulation.

\begin{figure*}
\centering
\includegraphics{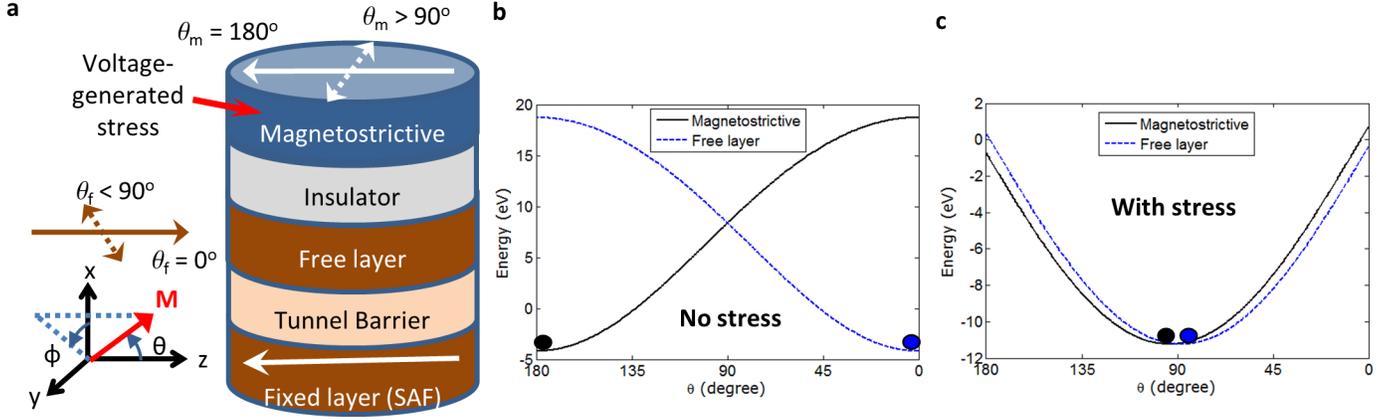}
\caption{\label{fig:potential_profile_dipole_coupled} {A magnetostrictive nanomagnet dipole coupled with a free layer of an MTJ with an effective field exerted from a fixed layer, and the potential landscapes of the magnetostrictive nanomagnet and the free layer without and with application of stress on the magnetostrictive nanomagnet.}
\textbf{(a)} A single-domain magnetostrictive nanomagnet lying on the $y$-$z$-plane. The magnetostrictive nanomagnet is dipole coupled with the free layer in the MTJ. The fixed layer of the MTJ is exerting an effective field $H_{fixed,f}$ on the free layer.
\textbf{(b)} The potential landscapes of the magnetostrictive nanomagnet and the free layer when no stress is applied on the magnetostrictive nanomagnet. Due to dipole coupling, the potential landscapes of both the nanomagnets are monostable. The magnetizations of the magnetostrictive nanomagnet and the free layer are pointing toward $\theta_m=180^\circ$ and $\theta_f=0^\circ$, respectively. An effective field of $H_{fixed,f}=0.1\,H_{k,f}$ is assumed to be exerted on the free layer, however, such small field has little effect on the potential landscapes.
\textbf{(c)} A stress of 30 MPa is applied on the Terfenol-D magnetostrictive nanomagnet to shift its minimum energy position around $\theta_m=90^\circ$. Due to dipole coupling, the free layer's magnetization also comes towards $\theta_f=90^\circ$. Note that $\theta_m > 90^\circ$ and $\theta_f < 90^\circ$, which is due to the exerted field of $H_{fixed,f}=0.1\,H_{k,f}$ on the free layer from the fixed layer.
}
\end{figure*}

Then we will consider a dipole-coupled structure between the magnetostrictive nanomagnet in a multiferroic composite and the free layer of an magnetic tunnel junction (see Fig.~\ref{fig:potential_profile_dipole_coupled}a). This is a more complicated and realistic design since read unit (magnetic tunnel junction) and write unit (multiferroic composite) are separated here so that the best materials for the magnetostrictive and free layer nanomagnets can be utilized.

\subsection{A magnetostrictive nanomagnet with an effective field from a fixed layer}

Figure~\ref{fig:potential_profile_magnetostrictive}a shows a single-domain magnetostrictive nanomagnet, on which an effective field is exerted from the fixed layer (synthetic antiferromagnetic (SAF) layer~\cite{RefWorks:300}). This is the case for which \emph{continuous} rotation of magnetization with stress is shown by solving stochastic LLG simulation.

We can write the total energy of the polycrystalline magnetostrictive nanomagnet as the sum of three energies -- the shape anisotropy energy, the stress anisotropy energy, and the anisotropy energy due to an effective field from the fixed layer in the $\theta=180^\circ$ direction (see Fig.~\ref{fig:potential_profile_magnetostrictive}a) -- as follows:

\begin{align}
E_{total}(\theta,\phi,t) &= E_{shape}(\theta,\phi) + E_{stress}(\theta,t) + E_{fixed}(\theta) \nonumber\\
												 &= B_{shape}(\phi) \, sin^2\theta - B_{stress}(t)\,cos^2\theta \nonumber\\
												 & \hspace*{4cm} + B_{fixed} \, cos\,\theta, \nonumber \displaybreak[3]\\
												 &= B(\phi) \, sin^2\theta + B_{fixed} \, cos\,\theta - B_{stress}(t)
\label{eq:total_anisotropy}
\end{align}
where
\begin{align}
B(\phi,t)&=B_{shape}(\phi)+B_{stress}(t),\displaybreak[3]\\
B_{shape}(\phi) &= (\mu_0/2) \, M_s \Omega \, [H_k  + H_d\, cos^2\phi],\displaybreak[3]\\
B_{stress}(t) &= (3/2) \, \lambda_s \sigma(t) \Omega,\displaybreak[3]\\
B_{fixed} &= \mu_0 M_s H_{fixed} \Omega,\displaybreak[3]
\label{eq:anisotropies}
\end{align}
\noindent
$M_s$ is the saturation magnetization, $H_k=(N_{d-yy}-N_{d-zz})\,M_s$ is the Stoner-Wohlfarth switching field~\cite{RefWorks:157}, $H_d=(N_{d-xx}-N_{d-yy})\,M_s$ is the out-of-plane demagnetization field~\cite{RefWorks:157}, $N_{d-mm}$ is the m$^{th}$ ($m=x,y,z$) component of the demagnetization factor, which depends on the dimensions of the nanomagnet~\cite{RefWorks:402} ($N_{dx} \gg N_{dy} > N_{dz}$), $\Omega$ is the volume of the magnetostrictive nanomagnet, $(3/2) \lambda_s$ is the magnetostriction coefficient of the magnetostrictive nanomagnet~\cite{RefWorks:157}, $\sigma(t)$ is the stress at time $t$, and $H_{fixed}$ is the effective field exerted by the fixed layer. 

Note that a negative $\lambda_s \sigma(t)$ product will favor alignment of the magnetization along the minor axis ($y$-axis) and according to our convention, in a material having \emph{positive} $\lambda_s$ (Terfenol-D i.e., TbDyFe, Galfenol i.e., FeGa), a sufficiently high \emph{compressive} stress that can overcome the in-plane shape-anisotropic barrier will rotate the magnetization toward the in-plane hard axis ($\theta=90^\circ$, $\phi=\pm90^\circ$). In Fig.~\ref{fig:potential_profile_magnetostrictive}c, note that the minimum energy position is not exactly $\theta=90^\circ$, which is due to the effective field $H_{fixed}$ exerted from the fixed layer on the magnetostrictive nanomagnet to make the potential landscape monostable (see Fig.~\ref{fig:potential_profile_magnetostrictive}b).

The magnetization \textbf{M} of the nanomagnet has a constant magnitude but a variable direction, so that we can represent it by a vector of unit norm $\mathbf{n_m} =\mathbf{M}/|\mathbf{M}| = \mathbf{\hat{e}_r}$ where $\mathbf{\hat{e}_r}$ is the unit vector in the radial direction in spherical coordinate system represented by ($r$,$\theta$,$\phi$). The other two unit vectors in the spherical coordinate system are denoted by $\mathbf{\hat{e}_\theta}$ and $\mathbf{\hat{e}_\phi}$ for $\theta$ and $\phi$ rotations, respectively. 

The effective field and torque acting on the magnetization due to gradient of potential landscape can be expressed as 

\begin{align}
\mathbf{H_{eff}}(\theta,\phi,t)  &= - \nabla E_{total}(\theta,\phi,t) \nonumber\\
									&= - \cfrac{\partial E_{total}}{\partial \theta}\,\mathbf{\hat{e}_{\theta}} - \cfrac{1}{sin\theta}\,\cfrac{\partial E_{total}}{\partial \phi}\,\mathbf{\hat{e}_\phi}, \\
	 \mathbf{T_{E}}(\theta,\phi,t) &= \mathbf{n_{m}} \times \mathbf{H_{eff}}(\theta,\phi,t) \nonumber\\
								  &= - [B(\phi,t)\, sin(2\theta) - B_{fixed}\,sin\theta] \,\mathbf{\hat{e}_\phi}  \nonumber\\
									& \quad - B_{shape,\phi}(\phi)\,sin\theta \,\mathbf{\hat{e}_\theta},
\label{eq:torque}
\end{align}
\noindent
where 
\begin{align}
B_{shape,\phi}(\phi) &= -\cfrac{\partial B_{shape}(\phi)}{\partial \phi} \nonumber\\
										 &=(\mu_0/2) \, M_s \Omega \, H_d \,  sin(2\phi). 
\label{eq:B_Bshape_phi}
\end{align}

The effect of random thermal fluctuations is incorporated via a random magnetic field 
\begin{equation}
\mathbf{h}(t)= h_x(t)\mathbf{\hat{e}_x} + h_y(t)\mathbf{\hat{e}_y} + h_z(t)\mathbf{\hat{e}_z},
\end{equation}
where $h_i(t)$ ($i=x,y,z$) are the three components of the random thermal field in Cartesian coordinates. We assume the properties of the random field $\mathbf{h}(t)$ as described in Ref.~\cite{RefWorks:186}. The random thermal field can be written as~\cite{RefWorks:186}
\begin{equation}
h_i(t) = \sqrt{\frac{2 \alpha kT}{|\gamma| M_V \Delta t}} \; G_{(0,1)}(t) \quad (i \in x,y,z)
\label{eq:ht}
\end{equation}
\noindent
where $\alpha$ is the dimensionless phenomenological Gilbert damping parameter, $\gamma$ is the gyromagnetic ratio for electrons, $M_V=\mu_0 M_s \Omega$, $1/\Delta t$ is proportional to the attempt frequency of the thermal field, $G_{(0,1)}(t)$ is a Gaussian distribution with zero mean and unit variance, $k$ is the Boltzmann constant, and $T$ is temperature.

The thermal field and corresponding torque acting on the magnetization can be written, respectively as 

\begin{align}
\mathbf{H_{TH}}(\theta,\phi,t) &= P_\theta(\theta,\phi,t)\,\mathbf{\hat{e}_\theta}+P_\phi(\theta,\phi,t)\,\mathbf{\hat{e}_\phi}, \\
\mathbf{T_{TH}}(\theta,\phi,t) &= \mathbf{n_m} \times \mathbf{H_{TH}}(\theta,\phi,t),
\end{align}
where
\begin{align}
P_\theta(\theta,\phi,t) &= M_V  \lbrack h_x(t)\,cos\theta\,cos\phi + h_y(t)\,cos\theta sin\phi \nonumber\\
												& \hspace*{3cm} - h_z(t)\,sin\theta \rbrack, \label{eq:thermal_parts_theta}\\
P_\phi(\theta,\phi,t) &= M_V  \lbrack h_y(t)\,cos\phi -h_x(t)\,sin\phi\rbrack.
\label{eq:thermal_parts_phi}
\end{align}
\noindent

The magnetization dynamics under the action of the torques $\mathbf{T_{E}}(t)$ and  $\mathbf{T_{TH}}(t)$ is described by the stochastic Landau-Lifshitz-Gilbert (LLG) equation~\cite{RefWorks:162,RefWorks:161,RefWorks:186} as follows.
\begin{equation}
\frac{d\mathbf{n_m}}{dt} - \alpha \left(\mathbf{n_m} \times \frac{d\mathbf{n_m}}{dt} \right) = -\frac{|\gamma|}{M} \left\lbrack \mathbf{T_E} +  \mathbf{T_{TH}}\right\rbrack. \label{eq:LLG}
\end{equation}

Solving the above equation analytically, we get the following coupled equations of magnetization dynamics for $\theta$ and $\phi$:
\begin{align}
\left(1+\alpha^2 \right) \frac{d\theta}{dt} &= \frac{|\gamma|}{M_V} \lbrack  - \alpha B(\phi,t) sin(2\theta) + \alpha B_{fixed}(t)\, sin\theta \nonumber \\
				& \qquad + B_{shape,\phi}(\phi) sin\theta \nonumber\\
				& \qquad + \left(\alpha P_\theta(\theta,\phi,t) + P_\phi (\theta,\phi,t) \right) \rbrack,
 \label{eq:theta_dynamics}
\end{align}
\begin{align}
\left(1+\alpha^2 \right) \frac{d \phi}{dt} &= \frac{|\gamma|}{M_V} \lbrack 2 B(\phi,t) cos\theta - B_{fixed}(t) \nonumber\\
					& \qquad + \alpha B_{shape,\phi}(\phi) \nonumber\\
					& \qquad - {[sin\theta]^{-1}} \left(P_\theta (\theta,\phi,t) - \alpha P_\phi (\theta,\phi,t) \right) \rbrack \nonumber \\
					& \hspace*{4.5cm} (sin\theta \neq 0).
\label{eq:phi_dynamics}
\end{align}
We need to solve the above two coupled equations numerically to track the trajectory of magnetization over time, in the presence of thermal fluctuations.

The energy dissipated in the nanomagnet due to Gilbert damping can be expressed as  $E_d = \int_0^{\tau}P_d(t) dt$, where $\tau$ is the time period when the input signal is active and $P_d(t)$ is the power dissipated at time $t$ per unit volume given by
\begin{equation}
P_d(t) = \frac{\alpha \, |\gamma|}{(1+\alpha^2) M_V} \, |\mathbf{T_E} (\theta(t), \phi(t), t)|^2 .
\label{eq:power_dissipation}
\end{equation}
Thermal field with mean zero does not cause any net energy dissipation but it causes variability in the energy dissipation by scuttling the trajectory of magnetization.

\emph{Initial fluctuation of magnetization.--} When the magnetization direction is {\it exactly} along the easy axis, i.e., $\sin\theta=0$ ($\theta=0^\circ$ or $\theta=180^\circ$), the torque acting on the magnetization given by Equation~\eqref{eq:torque} becomes zero. That is why only thermal fluctuations can deflect the magnetization vector \emph{exactly} from the easy axis. Considering the situation when $\theta=180^\circ$, from Equations~\eqref{eq:theta_dynamics} and~\eqref{eq:phi_dynamics}, we get~\cite{roy13_2}
\begin{equation}
\phi(t) = tan^{-1} \left( \frac{\alpha h_y(t) + h_x(t)}{\alpha h_x(t)-h_y(t)} \right),
\label{eq:phi_dynamics_thermal}
\end{equation}
\begin{equation}
\frac{d\theta(t)}{dt} =  \frac{-|\gamma| (h_x^2(t) + h_y^2(t))}{\sqrt{(\alpha h_x(t) - h_y(t))^2 + (\alpha h_y(t) + h_x(t))^2}}.
\label{eq:theta_dynamics_thermal}
\end{equation}
\noindent
From the Equation~\eqref{eq:theta_dynamics_thermal}, we notice clearly that the thermal fluctuations can deflect the magnetization \emph{exactly} from the easy axis since the time rate of change of $\theta(t)$ is non-zero in the presence of thermal fluctuations. Note that $d\theta(t)/dt$ does not depend on the component of the random thermal field along the $z$-axis, i.e. $h_z(t)$, which is a consequence of having $z$-axis as the easy axis of the nanomagnet. However, once the magnetization direction is even slightly deflected from the easy axis, all three components of the random thermal field along the $x$-, $y$-, and $z$-direction would come into play.

\subsection{A magnetostrictive nanomagnet magnetically coupled with a free layer with an effective field exerted from a fixed layer}

Figure~\ref{fig:potential_profile_dipole_coupled}a shows a magnetostrictive nanomagnet lying on the $y$-$z$ plane and it is dipole coupled with the free layer in an MTJ. The strong dipole coupling makes the potential landscapes of both the nanomagnets monostable as shown in the Fig.~\ref{fig:potential_profile_dipole_coupled}b. The fixed layer can exert an effective field on the free layer, however, that is not necessary to make the potential landscape of the magnetostrictive nanomagnet monostable, as required for the case in Fig.~\ref{fig:potential_profile_magnetostrictive}a. Such magnetically coupled design has been presented in Ref.~\cite{roy15_1} in the context of separating read and write units for digital computing purposes, however, as depicted in the Fig.~\ref{fig:roy_fig1}, there is difference in the potential landscapes of the nanomagnets for analog computing.

The magnetization $\mathbf{M_m}$ ($\mathbf{M_f}$) of the magnetostrictive (free layer) nanomagnet has a constant magnitude  but a variable direction, so that we can represent it by a vector of unit norm $\mathbf{m_{m}} =\mathbf{M_m}/|\mathbf{M_m}| = \mathbf{\hat{e}_{r}}$ ($\mathbf{m_{f}} =\mathbf{M_f}/|\mathbf{M_f}| = \mathbf{\hat{e}_{r}}$) where $\mathbf{\hat{e}_{r}}$ is the unit vector in the radial direction in spherical coordinate system represented by ($r$,$\theta$,$\phi$). The other two unit vectors corresponding to the polar angle $\theta$ and the azimuthal angle $\phi$ are $\mathbf{\hat{e}_{\theta}}$ and $\mathbf{\hat{e}_{\phi}}$, respectively. We will use the subscripts $m$ and $f$ to denote any parameter for the \emph{magnetostrictive} nanomagnet and the \emph{free layer} nanomagnet, respectively. 

While there are two nanomagnets (magnetostrictive and free layer) having a dipole coupling between them (see Fig.~\ref{fig:potential_profile_dipole_coupled}a), stress is generated only on the magnetostrictive nanomagnet. Hence we have additional stress anisotropy to consider for the magnetostrictive nanomagnet. Also, we consider that the fixed layer in general exerts an effective field on the free layer. The potential energies of the single-domain polycrystalline \emph{magnetostrictive} nanomagnet and the \emph{free layer} nanomagnet can be expressed, respectively, as
\begin{align}
& E_{total,m} (\theta_m,\phi_m,\theta_f,\phi_f,t) \nonumber \\
 & \, = E_{shape,m}(\theta_m,\phi_m) + E_{stress}(t) + E_{dipole}(\theta_m,\phi_m,\theta_f,\phi_f) \nonumber\\
 & \, = B_m(\phi_m,t) \, sin^2 \theta_m + E_{dipole}(\theta_m,\phi_m,\theta_f,\phi_f),
\label{eq:total_anisotropy_magnetostrictive}
\end{align}
and
\begin{align}
& E_{total,f} (\theta_m,\phi_m,\theta_f,\phi_f) \nonumber \\
& \, = E_{shape,f}(\theta_f,\phi_f) + E_{fixed}(\theta_f) + E_{dipole}(\theta_m,\phi_m,\theta_f,\phi_f) \nonumber\\
& \, = B_f(\phi_f) \, sin^2 \theta_f - B_{fixed}cos\,\theta_f + E_{dipole}(\theta_m,\phi_m,\theta_f,\phi_f),
\label{eq:total_anisotropy_free_layer}
\end{align}
where
\begin{align}
B_m(\phi_m,t) &= B_{shape,m}(\phi_m) + B_{stress}(t),\displaybreak[3]\\
B_{shape,m}(\phi_m) &= (\mu_0/2) M_{s,m} \Omega_m [H_{k,m} + H_{d,m} \,cos^2\phi_m],\\
B_{stress}(t) 	&= (3/2) \lambda_s \sigma (t) \Omega_m,
\label{eq:shape_stress_magnetostrictive}
\end{align}
\begin{align}
B_f(\phi_f) &= B_{shape,f}(\phi_f),\\
B_{fixed} &= \mu_0 M_{s,f} \Omega_f H_{fixed,f},\\
B_{shape,f}(\phi_f) &= (\mu_0/2) M_{s,f} \Omega_f [H_{k,f} + H_{d,f}\,cos^2\phi_f],
\label{eq:shape_free_layer}
\end{align}
\begin{align}
	& E_{dipole}(\theta_m,\phi_m,\theta_f,\phi_f) \nonumber\\
	& \hspace*{2mm} = \cfrac{\mu_0}{4\pi R^3} \, M_{s,m} \Omega_m M_{s,f} \Omega_f \lbrack cos\theta_m cos\theta_f \nonumber \\
	& \hspace*{2mm} \quad + sin \theta_m sin \theta_f (sin \phi_m sin\phi_f - 2 cos\phi_m cos\phi_f) \rbrack \displaybreak[3],
\label{eq:dipole}	
\end{align}
where
$M_{s,m(f)}$ is the saturation magnetization of the magnetostrictive (free layer) nanomagnet, $H_{k,m(f)}=(N_{d-yy,m(f)}-N_{d-zz,m(f)})\,M_{s,m(f)}$ is the Stoner-Wohlfarth switching field~\cite{RefWorks:157} for the magnetostrictive (free layer) nanomagnet, $H_{d,m(f)}=(N_{d-xx,m(f)}-N_{d-yy,m(f)})\,M_{s,m(f)}$ is the out-of-plane demagnetization field~\cite{RefWorks:157} for the magnetostrictive (free layer) nanomagnet, $N_{d-pp,m(f)}$ is the component of the demagnetization factor for the magnetostrictive (free layer) nanomagnet along $p$-direction, which depends on the nanomagnet's dimensions~\cite{RefWorks:157,RefWorks:402}, $\Omega_m$ ($\Omega_f$) is the volume of the magnetostrictive (free layer) nanomagnet, $(3/2)\lambda_s$ is the magnetostrictive coefficient of the single-domain magnetostrictive nanomagnet~\cite{RefWorks:157}, $\sigma(t)$ is the stress on the magnetostrictive nanomagnet at time $t$, and $H_{fixed,f}$ is the effective field exerted by the fixed layer on the free layer, and $\mathbf{R}=R\,{\mathbf{\hat{e}_{x}}}$ is the center-to-center distance between the nanomagnets. 

Note that the product of magnetostrictive coefficient and stress needs to be \emph{negative} in sign so that the generated stress-anisotropy can beat the shape-anisotropy to rotate the magnetization of the magnetostrictive nanomagnet. As the magnetization of the magnetostrictive nanomagnet rotates upon application of stress, the magnetization of the free layer nanomagnet does also rotate due to the dipole coupling between the nanomagnets. An effective field is exerted from the fixed layer on the free layer to keep the respective orientations of the magnetostrictive nanomagnet and the free layer intact, i.e., when stress is removed, the magnetostrictive nanomagnet should be pointing to $\theta_f=0^\circ$ and free layer's magnetization should be pointing to $\theta_m=180^\circ$ (see Fig.~\ref{fig:potential_profile_dipole_coupled}b).

We can write the LLG equations for the individual nanomagnets following the Equation~\eqref{eq:LLG}, and  we will also use the corresponding terms in the Equations~\eqref{eq:thermal_parts_theta} and~\eqref{eq:thermal_parts_phi} $P_{\theta_{m(f)}}$ and $P_{\phi_{m(f)}}$ for the magnetostrictive (free layer) nanomagnet, $\alpha_{m(f)}$ as the phenomenological damping parameter of the magnetostrictive (free layer) nanomagnet, and $M_{V,m(f)}= \mu_0 M_{s,m(f)} \Omega_{m(f)}$. After solving the two LLG equations, we get the following coupled equations for the dynamics of $\theta_m$ and $\phi_m$:
\begin{multline}
\left(1+\alpha_m^2 \right) \cfrac{d\theta_m}{dt} = \cfrac{|\gamma|}{M_{V,m}} [ B_{shape,\phi_m}(\phi_m)sin\theta_m \\
 - 2\alpha_m B_m(\phi_m) sin\theta_m cos\theta_m - T_{dipole,\theta_m} - \alpha_m T_{dipole,\phi_m} \\
 + (\alpha_m P_{\theta_m} + P_{\phi_m})],
 \label{eq:theta_dynamics_magnetostrictive}
\end{multline}
\begin{multline}
\left(1+\alpha_m^2 \right) \cfrac{d\phi_m}{dt} = \cfrac{|\gamma|}{M_{V,m}} \cfrac{1}{sin\theta_m} [\alpha_m B_{shape,\phi_m}(\phi_m)sin\theta_m \\
	 + 2 B_m(\phi_m) sin\theta_m cos\theta_m + \alpha_m T_{dipole,\theta_m} + T_{dipole,\phi_m} \\
	 - \{sin\theta_m\}^{-1} (P_{\theta_m} - \alpha_m P_{\phi_m})] \qquad (sin\theta_m \neq 0),
  \label{eq:phi_dynamics_magnetostrictive}
\end{multline}
and the following coupled equations for the dynamics of $\theta_f$ and $\phi_f$:
\begin{multline}
\left(1+\alpha_f^2 \right) \cfrac{d\theta_f}{dt} = \cfrac{|\gamma|}{M_{V,f}} [ B_{shape,\phi_f}(\phi_f)sin\theta_f \\
 - 2\alpha_f B_f(\phi_f) sin\theta_f cos\theta_f - \alpha_f B_{fixed} sin(\theta_f) \\
 - T_{dipole,\theta_f} - \alpha_f T_{dipole,\phi_f} + (\alpha_f P_{\theta_f} + P_{\phi_f})],
 \label{eq:theta_dynamics_free_layer}
\end{multline}
\begin{multline}
\left(1+\alpha_f^2 \right) \cfrac{d\phi_f}{dt} = \cfrac{|\gamma|}{M_{V,f}} \cfrac{1}{sin\theta_f} [\alpha_f B_{shape,\phi_f}(\phi_f)sin\theta_f \\
	 + 2 B_f(\phi_f) sin\theta_f cos\theta_f  + B_{fixed} \\
	 + \alpha_f T_{dipole,\theta_f} + T_{dipole,\phi_f} - \{sin\theta_f\}^{-1} (P_{\theta_f} - \alpha_f P_{\phi_f})] \\
	  (sin\theta_f \neq 0),
  \label{eq:phi_dynamics_free_layer}
\end{multline}
where
\begin{align}
& B_{shape,\phi_{m(f)}}(\phi_{m(f)}) = -\cfrac{\partial B_{shape,m(f)}(\phi_{m(f)})}{\partial \phi_{m(f)}} \nonumber\\
												 & \qquad= (\mu_0/2) \, M_{s,m(f)} \Omega_{m(f)} H_{d,m(f)} sin(2\phi_{m(f)}),
\label{eq:B_shape_phi_magnetostrictive}
\end{align}
\begin{align}
T_{dipole,\theta_{m(f)}} &= \cfrac{1}{sin \theta_{m(f)}}\,\cfrac{\partial E_{dipole}}{\partial \phi_{m(f)}},\\
T_{dipole,\phi_{m(f)}} &= \cfrac{\partial E_{dipole}}{\partial \theta_{m(f)}}. 
\end{align}
The magnetization dynamics of the two nanomagnets represented by the Equations~\eqref{eq:theta_dynamics_magnetostrictive},~\eqref{eq:phi_dynamics_magnetostrictive},~\eqref{eq:theta_dynamics_free_layer}, and~\eqref{eq:phi_dynamics_free_layer}, are coupled through the dipole coupling [see Equation~\eqref{eq:dipole}]. These coupled equations are solved numerically to track the trajectories of the two magnetizations over time.

The individual internal energy dissipation in the magnetostrictive nanomagnet and the free layer due to Gilbert damping $E_{d,m}$ and $E_{d,f}$, respectively can be expressed following the prescription in Equation~\eqref{eq:power_dissipation}. We sum up these two internal energy dissipations $E_{d,m}$ and $E_{d,f}$ to get the total energy dissipation $E_d$ in the nanomagnets. Since the magnetizations of the two magnetically coupled nanomagnets rotate quite concomitantly, a single time period $\tau$ when the input signal is active is sufficient. 

\emph{Initial fluctuations of the magnetizations.--} When the magnetization direction of the magnetostrictive nanomagnet and the free layer's magnetization are \emph{exactly} along $\theta_f=0^\circ$ and $\theta_m=180^\circ$, respectively (Fig.~\ref{fig:potential_profile_dipole_coupled}b), the torque acting on the magnetizations become zero. That is why only thermal fluctuations can deflect the magnetization vector from this lockjam. For the magnetostrictive nanomagnet, when $\theta_m=180^\circ$, we can derive the following:
\begin{equation}
\phi_m(t) = tan^{-1} \left( \cfrac{\alpha h_{y,m}(t) + h_{x,m}(t)}{\alpha h_{x,m}(t)-h_{y,m}(t)} \right),
\label{eq:phi_dynamics_inital_magnetostrictive}
\end{equation}
\begin{equation}
\frac{d\theta_m}{dt} =  \cfrac{-|\gamma| (h_{x,m}^2(t) + h_{y,m}^2(t))}{\sqrt{(\alpha h_{x,m}(t) - h_{y,m}(t))^2 + (\alpha h_{y,m}(t) + h_{x,m}(t))^2}}.
\label{eq:theta_dynamics_inital_magnetostrictive}
\end{equation}
\noindent
Similarly, for the free layer, when $\theta_f=0^\circ$, we can derive the following:
\begin{equation}
\phi_f(t) = tan^{-1} \left( \cfrac{\alpha h_{y,f}(t) - h_{x,f}(t)}{\alpha h_{x,f}(t) + h_{y,f}(t)} \right),
\label{eq:phi_dynamics_inital_free}
\end{equation}
\begin{equation}
\frac{d\theta_f}{dt} =  \cfrac{|\gamma| (h_{x,f}^2(t) + h_{y,f}^2(t))}{\sqrt{(\alpha h_{x,f}(t) - h_{y,f}(t))^2 + (\alpha h_{y,f}(t) + h_{x,f}(t))^2}}.
\label{eq:theta_dynamics_inital_free}
\end{equation}
From the Equations~\eqref{eq:theta_dynamics_inital_magnetostrictive} and~\eqref{eq:theta_dynamics_inital_free}, we notice clearly that the thermal fluctuations can deflect the magnetizations \emph{exactly} from their respective easy axis since the time rate of change of $\theta_m(t)$ and $\theta_f(t)$ is non-zero in the presence of thermal fluctuations. The sign of the rates of change of $\theta_m(t)$ and $\theta_f(t)$ are in the opposite direction, which signify that $\theta_m$ decreases from $180^\circ$ and $\theta_f$ increases from $0^\circ$ with random thermal fluctuations, facilitating the rotation of the magnetically coupled magnetizations. Note that $d\theta_{m(f)}(t)/dt$ does not depend on the component of the random thermal field along the $z$-axis, which is a consequence of having $z$-axis as the easy axis of the nanomagnet. However, once the magnetization direction is even slightly deflected, all three components of the random thermal field along the $x$-, $y$-, and $z$-direction would come into play.

\vspace*{2mm}

It should be pointed out that the orientations of the magnetostrictive nanomagnet and the free layer are not \emph{exactly} along $\theta_m=180^\circ$ and $\theta_f=0^\circ$, due to thermal fluctuations. The mean deflection of the magnetizations due to room-temperature (300 K) thermal fluctuations, when no stress is applied, is $\sim 3.25^\circ$. This is a bit lower than that of the case for Fig.~\ref{fig:potential_profile_magnetostrictive}b (mean deflection of magnetization around $\theta=180^\circ$ is $\sim 3.9^\circ$), due to the strong dipole coupling between the magnetostrictive nanomagnet and the free layer in Fig~\ref{fig:potential_profile_dipole_coupled} for the same dimensions of the nanomagnets. On the contrary, such strong magnetic dipole coupling can help reducing the dimensions of the nanomagnets.

\section{Material parameters and device dimensions}
\label{sec:mat_params_dimensions}

The magnetostrictive nanomagnet is made of polycrystalline Terfenol-D and it has the following material properties -- saturation magnetization ($M_{s,m}$):  8$\times$10$^5$ A/m, Gilbert damping parameter ($\alpha_m$): 0.1, Young's modulus (Y): 80 GPa, magnetostrictive coefficient ($(3/2)\lambda_s$): +90$\times$10$^{-5}$, and Poisson's ratio ($\nu$): 0.3 (Refs.~\cite{RefWorks:179,RefWorks:176,RefWorks:178,RefWorks:821,roy13_spin,roy11_6,roy13_2,roy11_news,roy11_2,roy13,roy14}) For the magnetically-coupled structure (Fig.~\ref{fig:potential_profile_dipole_coupled}a), the free layer nanomagnet is made of widely used CoFeB, which has the following material properties -- Gilbert damping parameter ($\alpha_f$): 0.01, saturation magnetization ($M_{s,f}$):  8$\times$10$^5$ A/m~\cite{RefWorks:413}. 

For the piezoelectric layer, we use lead magnesium niobate-lead titanate (PMN-PT), which has a dielectric constant of 1000, $d_{31}$=--3000 pm/V, and $d_{32}$=1000 pm/V~\cite{RefWorks:790}. The piezoelectric layer's thickness $t_{piezo}$=50 nm (similar dimension is used in Ref.~\cite{roy11_6}). Using low-thickness piezoelectric layers [e.g., $<$ 100 nm of PMN-PT~\cite{RefWorks:820}] avoiding any substrate clamping effect~\cite{RefWorks:823,RefWorks:820,RefWorks:170} requires further experimental efforts. There are other multiferroic composites, e.g., Ni/PMN-PT~\cite{RefWorks:836,RefWorks:551,RefWorks:611}, FeGaB/PMN-PT~\cite{RefWorks:801}, CoFe/PMN-PT~\cite{RefWorks:790}, CoFeB/PMN-PT~\cite{RefWorks:863,RefWorks:838,RefWorks:520}, $Fe_3O_4$/PMN-PT~\cite{RefWorks:827}, $CoFe_2O_4$/PMN-PT~\cite{RefWorks:830,RefWorks:839}, however, Terfenol-D/PZN-PT~\cite{RefWorks:806} or Terfenol-D/PMN-PT~\cite{pmnpt} composites possess superior material parameters leading to high magnetoelectric coupling. FeGaB/PMN-PT composite has also high magnetoelectric couplings similar to Terfenol-D/PMN-PT. 

For the magnetic tunnel junction in the magnetically-coupled structure (Fig.~\ref{fig:potential_profile_dipole_coupled}a), the resistance-area (RA) product (for parallel orientation), and tunneling magnetoresistance (TMR) are assumed to be 175 k$\Omega$-$\mu m^2$  and 300\%~\cite{RefWorks:33}, respectively. MTJs with RA product as low as 4.3 $\Omega$-$\mu m^2$ (Ref.~\cite{RefWorks:786}) and 1000\% TMR with half-metals~\cite{RefWorks:581} are plausible. 

The dimensions of the single-domain nanomagnets are chosen as $100\,nm \times 90\,nm \times 6\,nm$,~\cite{RefWorks:402,RefWorks:133} and the center-to-center distance between the nanomagnets in the magnetically-coupled structure (Fig.~\ref{fig:potential_profile_dipole_coupled}a) is $R=40\,nm$. We choose the dimensions of the nanomagnet such that it has always a single ferromagnetic domain~\cite{RefWorks:402,RefWorks:133}. (At higher dimensions, there are issues like inhomogeneous magnetization, buckling instability, nonuniform magnetization reversal, vortex dynamics etc.~\cite{RefWorks:730,RefWorks:731,RefWorks:733,RefWorks:732,RefWorks:734,RefWorks:735}.) This allows us to use the so-called \emph{macrospin} model, which is valid at low dimensions~\cite{RefWorks:422,RefWorks:105,RefWorks:7}, as chosen here.

\begin{figure*}
\centering
\includegraphics[width=\textwidth]{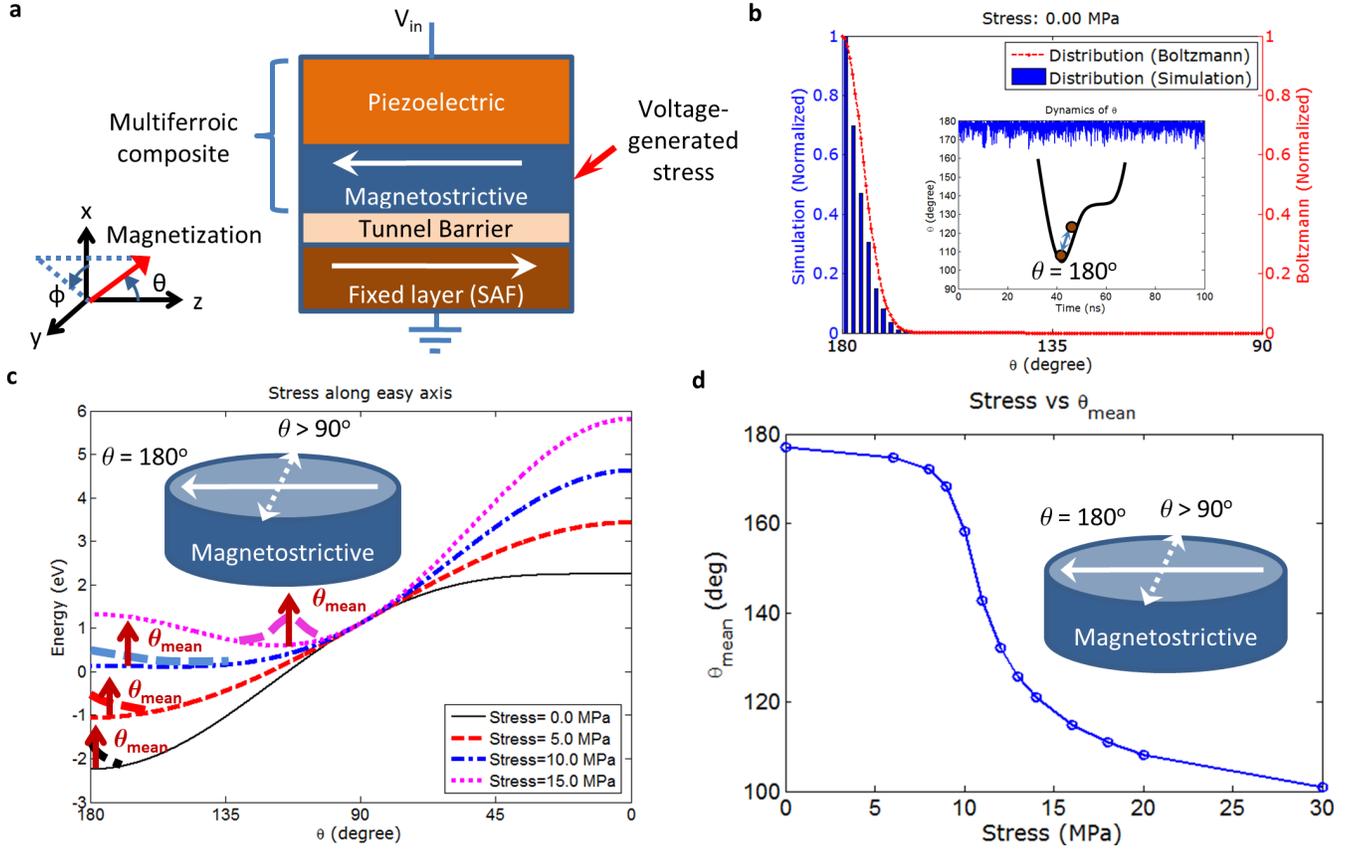}
\caption{\label{fig:roy_fig2} {Continuous rotation of magnetization for analog signal processing using strain-mediated multiferroic composites.}
\textbf{(a)} The device structure consists of a multiferroic composite (piezoelectric layer strain-coupled with a magnetostrictive free layer) attached to a tunnel barrier (e.g., MgO) and a fixed nanomagnet for tunneling magnetoresistance (TMR) measurement. The contact electrodes are not shown explicitly and the vertical structure can be inverted, i.e., the piezoelectric layer can be at bottom too.
\textbf{(b)} Magnetization is fluctuating due to room temperature thermal fluctuations at one potential well when no stress is applied. This accords to the Boltzmann distribution signifying the validity of our model, which is based on stochastic Landau-Lifshitz-Gilbert equation of magnetization dynamics at room-temperature (300 K). The inset shows the the fluctuation of magnetization in time-domain.
\textbf{(c)} As stress modulates the potential landscape of the nanomagnet, thermal fluctuations create a distribution of magnetization. For 0, 5 MPa, and 10 MPa stresses, the potential minima is at $\theta=180^\circ$, however, the mean value of magnetization's distribution varies \emph{continuously}. When the stress is high enough, potential minimum shifts towards $\theta=90^\circ$.
\textbf{(d)} Stress versus mean orientation of magnetization of the magnetostrictive nanomagnet. This signifies a \emph{continuous} rotation of magnetization with a high-slope region, as required for analog signal processing.
} 
\end{figure*}

\section{Simulation Results and Discussions}
\label{sec:results}

\begin{figure*}
\centering
\includegraphics[width=\textwidth]{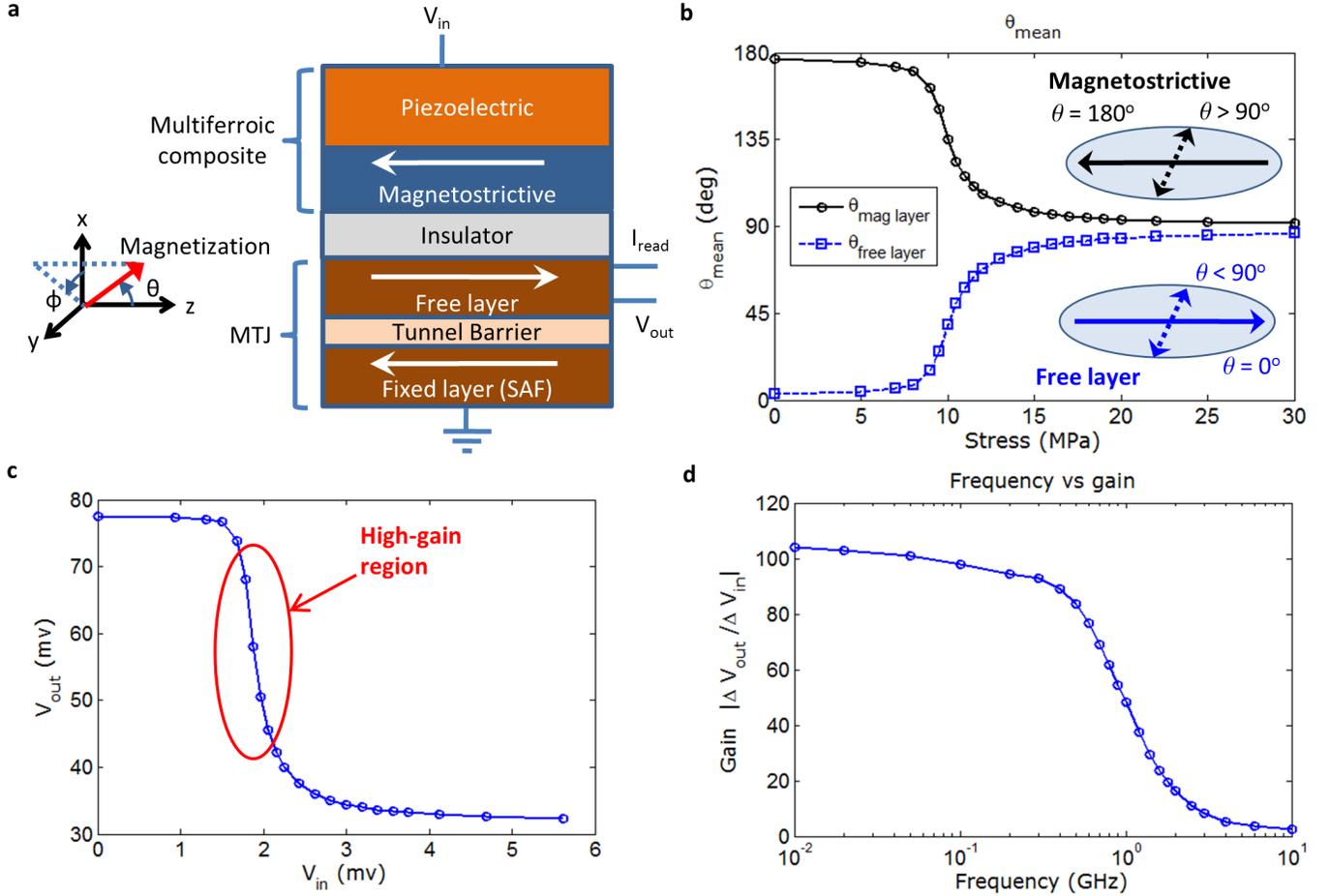}
\caption{\label{fig:roy_fig3} {An amplifier characteristics using strain-mediated multiferroic composites.} 
\textbf{(a)} A multiferroic composite and a magnetic tunnel junction (MTJ) separated by an insulator. The output of the device is extracted from the TMR measurement of the MTJ structure by passing a read current $I_{read}$, which reads the state of the free layer and in turn the state of the magnetostrictive layer since they form antiparallel orientation due to dipole coupling. The contact electrodes are not shown explicitly and the vertical structure can be inverted, i.e., the piezoelectric layer can be at bottom too. Note that the read current flows through the MTJ only and not through the piezoelectric layer.
\textbf{(b)} Simulation results using stochastic Landau-Lifshitz-Gilbert (LLG) equation of magnetization dynamics to show that the magnetostrictive layer and the free layer indeed rotate concomitantly due to dipole coupling when stress acts on the magnetostrictive layer. 
\textbf{(c)} Input voltage $V_{in}$ versus output voltage $V_{out}$ characteristics showing a high gain transition region where we can bias an input voltage and can achieve an amplified version of the input voltage at the output. 
\textbf{(d)} Frequency dependence of gain. Gain reduces as we increase the frequency of the input signal. The results show that a gain of 100 can be achieved at low frequency ($<$ 500 MHz), while a gain of 50 can be achieved at 1 GHz frequency.
}
\end{figure*}

Figure~\ref{fig:roy_fig2}a shows a multiferroic composite (piezoelectric layer strain-coupled with a magnetostrictive nanomagnet) and a fixed magnetization layer separated by a tunnel barrier e.g., Magnesium Oxide (MgO)~\cite{RefWorks:76,RefWorks:572,RefWorks:33,RefWorks:74}. An input voltage $V_{in}$ generates strain in the piezoelectric layer and the strain is elastically transferred to the magnetostrictive nanomagnet generating a stress anisotropy in it. This additional anisotropy can rotate the magnetization of the magnetostrictive nanomagnet~\cite{roy13_spin}. The rotation of magnetization can be detected using tunneling magnetoresistance measurement (in magnetic tunnel junctions (MTJs)~\cite{RefWorks:300}) with respect to the magnetization of the fixed layer. The fixed layer is implemented as a synthetic antiferromagnetic (SAF) layer~\cite{RefWorks:300}, which can be designed in a way to exert a field on the magnetostrictive nanomagnet to make the potential landscape monostable as shown in the Fig.~\ref{fig:roy_fig1}c. A magnetically coupled design between the multiferroic composite (write unit) and the MTJ (read unit) separating the write and read units~\cite{roy15_1} will be presented in the Fig.~\ref{fig:roy_fig3}a later. The layers are shown as rectangular slabs, but they could be of any anisotropic shape like elliptical cylinders as we have used here. 

Figure~\ref{fig:roy_fig2}b shows the room-temperature (300 K) distribution of magnetization (polar angle $\theta$) since magnetization is fluctuating due to thermal agitations around $\theta=180^\circ$. The model utilized to determine this distribution is the stochastic Landau-Lifshitz-Gilbert (LLG) equation~\cite{RefWorks:186} of magnetization dynamics as described earlier 
in the presence of thermal fluctuations (incorporated as a random thermal field in the LLG equation) and it matches the Boltzmann distribution (see Fig.~\ref{fig:roy_fig2}b) signifying the validity of our model. As voltage generated compressive stress is built on the magnetostrictive nanomagnet, it modifies the potential landscape of the nanomagnet and the mean orientation of magnetization ($\theta_{mean}$) varies \emph{continuously} due to thermal fluctuations as depicted in the Figs.~\ref{fig:roy_fig2}c and~\ref{fig:roy_fig2}d. When a sufficient compressive stress is applied, the magnetization comes near $\theta=90^\circ$. A few distributions for different stresses calculated from stochastic LLG are shown in the Fig.~\ref{fig:results_magnetostrictive} later.  

Note that the azimuthal angle $\phi$ can be either $+90^\circ$ or $-90^\circ$ (both correspond to magnet's plane) as depicted by the bidirectional arrow in Fig.~\ref{fig:roy_fig2}c when $\theta$ reaches around $90^\circ$. The azimuthal angle $\phi$ is imperative to consider since stress rotates magnetization out of magnet's plane and a slight out-of-plane excursion can create a high torque on the magnetization due to high out-of-plane demagnetization field of the thin nanomagnet leading to an increase of switching speed by a couple of orders in magnitude to sub-nanosecond~\cite{roy13_2}. In general the azimuthal angle $\phi$ has also a distribution due to thermal fluctuations~\cite{roy13_2}, however, due to the effective field exerted by the fixed layer on the magnetostrictive nanomagnet making the potential landscape monostable, the $\phi$-distribution is quite quenched and it facilitates the match with the Boltzmann distribution in Fig.~\ref{fig:roy_fig2}b considering only the $\theta$-space.

In Fig.~\ref{fig:roy_fig3}a, compared to the Fig.~\ref{fig:roy_fig2}a, we have introduced a dipole coupled free layer so that suitable materials can be utilized for both the magnetostrictive layer of the multiferroic composite and the free layer of the MTJ \emph{independently}. For the magnetostrictive layer, materials with high magnetostriction coefficient (Terfenol-D, Galfenol, i.e., FeGa) should be used, while for the free layer, relevant materials e.g., CoFeB, half-metals need to be used to achieve a high tunneling magnetoresistance (TMR)~\cite{RefWorks:76,RefWorks:572,RefWorks:33,RefWorks:74,RefWorks:581}. Although multiferroic composites employing CoFe~\cite{RefWorks:790,RefWorks:838} is possible, with the advent of half-metal based MTJs~\cite{RefWorks:581} leading to even higher TMR than that of using CoFeB and the requirement of using materials with higher magnetostriction coefficient than that of CoFeB leading to a lower operating voltage (and therefore a lower energy dissipation), clearly establishes the requirement to separate the read and write units. Note that the insulator does not aid in input-output isolation for logic design purposes, since the insulating piezoelectric layer with much higher resistivity makes the input-output isolation inherent in these multiferroic devices~\cite{roy13}. Rather than using dipolar coupling, exchange-coupled magnets can be also utilized, as mentioned in Ref.~\cite{roy15_1}.

Figure~\ref{fig:roy_fig3}b depicts that the magnetizations of the magnetostrictive nanomagnet and the free layer rotate concomitantly upon application of stress on the magnetostrictive nanomagnet. The model for this simulation has been described in the Section~\ref{sec:stochastic_LLG}. Since the design of the fixed layer (a synthetic antiferromagnetic layer) may not be very precise, an effective field of 10\% of the free layer's anisotropy field $H_{k,f}$ is assumed to be exerted on the free layer. (It also aids to keep the initial orientations of the magnetizations of the magnetostrictive nanomagnet and the free layer, when stress is withdrawn from the magnetostrictive nanomagnet.) Therefore, when a sufficient high stress is applied, the magnetization of the magnetostrictive nanomagnet comes to $94^\circ$ and not to exactly $90^\circ$. At around $\theta=90^\circ$, the magnetizations of the magnetostrictive nanomagnet and the free layer orient anti-parallel in $\phi$-space. Note that the insulator thickness separating the magnetostrictive nanomagnet and the free layer is optimized since at higher thicknesses the magnetizations do not quite rotate concomitantly and at lower thicknesses, the high coupling rotates magnetizations out-of-plane leading to precessional motion~\cite{roy15_1}.

\begin{figure}
\centering
\includegraphics{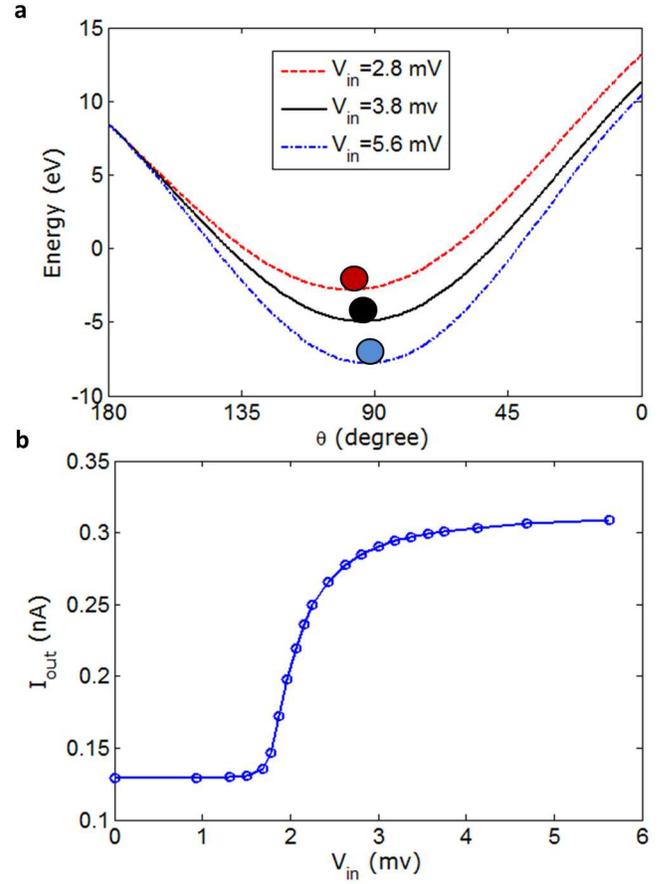}
\caption{\label{fig:roy_fig4} {A current source characteristics using strain-mediated multiferroic composites.} 
\textbf{(a)} The potential landscapes of the magnetostrictive nanomagnet for different voltages, when the voltages are sufficient enough to make the potential landscape monostable around $\theta=90^\circ$. The well goes deeper with the increase of voltages.
\textbf{(b)} Stochastic LLG simulation results for the current source characteristics. The characteristics show that as we increase the voltage over a certain value $V_{in} > 3$ mV, the current from the device becomes constant.
}
\end{figure}

Figure~\ref{fig:roy_fig3}c shows the input-output characteristics of the amplifier. The Appendices~\ref{sec:theta_vout} and~\ref{sec:voltage_stress_anisotropy} describe the MTJ characteristics to extract the magnetization orientation versus output voltage relationship and the input voltage versus stress relationship, respectively. This input-output characteristics is similar to the transistor input-output characteristics having a high-gain region. This characteristics considers 10\% of the free layer's anisotropy field $H_{k,f}$ to be exerted on the free layer by the fixed layer, however, the gain remains similar when no field is exerted from the fixed layer. If the exerted field from the fixed layer increases significantly, the gain of the amplifier reduces (see Fig.~\ref{fig:results_dipole_coupled_Hz} later). No effective field is considered to be exerted on the magnetostrictive layer by the fixed layer since it is 40 nm away~\cite{roy15_1}. The amplifier characteristics shown in the Fig.~\ref{fig:roy_fig3}c signify that both the DC level enhancement and AC signal gain can be achieved \emph{simultaneously} and the MTJ characteristics (see Fig.~\ref{fig:results_MTJ}) aids in achieving it. The output voltage level can be adjusted by changing the read current $I_{read}$ and resistance-area product of the MTJ (by changing the tunnel barrier thickness)~\cite{RefWorks:33,RefWorks:74}. To obtain a current source, the structure in Fig.~\ref{fig:roy_fig3}a can be utilized, while biasing the MTJ using a constant voltage source (see Fig.~\ref{fig:roy_fig4}).
Note that the dipole-coupled design in the Fig.~\ref{fig:roy_fig3}a achieves much better linearity in the high gain region compared to the case in the Fig.~\ref{fig:roy_fig2}a. Fig.~\ref{fig:roy_fig3}d plots the frequency dependence of gain. An AC voltage of different frequencies is applied at the input and the output voltage is extracted to determine the gain. This characteristics is similar to the transistor characteristics that at high frequency the gain reduces~\cite{razav01}. See Fig.~\ref{fig:results_dipole_coupled_input_output} later for detailed simulations in this regard.

The energy dissipation for the amplifier presented in the Fig.~\ref{fig:roy_fig3} turns out to be $\sim$0.1 attojoule/cycle at 1 GHz frequency (see Appendix~\ref{sec:energy_dissipation}). 
The experimental efforts on these multiferroic devices have demonstrated the existence of stress anisotropy and the stochastic LLG equation of magnetization dynamics used here is a standard tool to benchmark experimental results. Utilizing piezoelectric layers of low thickness ($<$ 100 nm) avoiding any substrate clamping effect, which needs to be tackled~\cite{RefWorks:820,RefWorks:823}, while retaining high piezoelectric coefficients will allow us to operate these devices at few millivolts incurring miniscule amount of energy dissipation and therefore possibly by energy harvesting~\cite{roundyf}. This is about 2-3 orders of magnitude less energy dissipation than the traditional charge based transistors operating at $\sim$1 Volt. Also, these multiferroic devices have advantage that these are voltage-controlled devices rather than current-controlled and the presence of thick piezoelectric layer leads to a miniscule leakage unlike the case of traditional transistors. The area and the frequency of operation for these devices can be further improved by using interface and exchange coupled systems, perpendicular anisotropy~\cite{roy14_2,RefWorks:815} etc. 

Figure~\ref{fig:roy_fig4} shows a characteristics of a current source (e.g., $I_{read}$ in Fig.~\ref{fig:roy_fig3}a) using the strain-mediated multiferroic devices. Fig.~\ref{fig:roy_fig4}a depicts the operating principle of building the current source. When sufficiently high input voltages are applied, the potential landscapes go deeper around $\theta=90^\circ$ and magnetization comes close to $\theta=90^\circ$. To obtain a current source, the structure in Fig.~\ref{fig:roy_fig3} can be utilized, while biasing the MTJ using a constant voltage source $V_{read}$. Fig.~\ref{fig:roy_fig4}b shows the stochastic LLG simulation results of the current source using $V_{read}=10$ mV. This is similar to the transistor characteristics driven in saturation that provides almost constant currents~\cite{razav01}. It should be mentioned that we need to consider the characteristics of the current source while connecting to a load depending on the load resistance. The input resistance of the current source can be increased by increasing the resistance-area product or the area of the MTJ for the current source.

\begin{figure}
\centering
\includegraphics{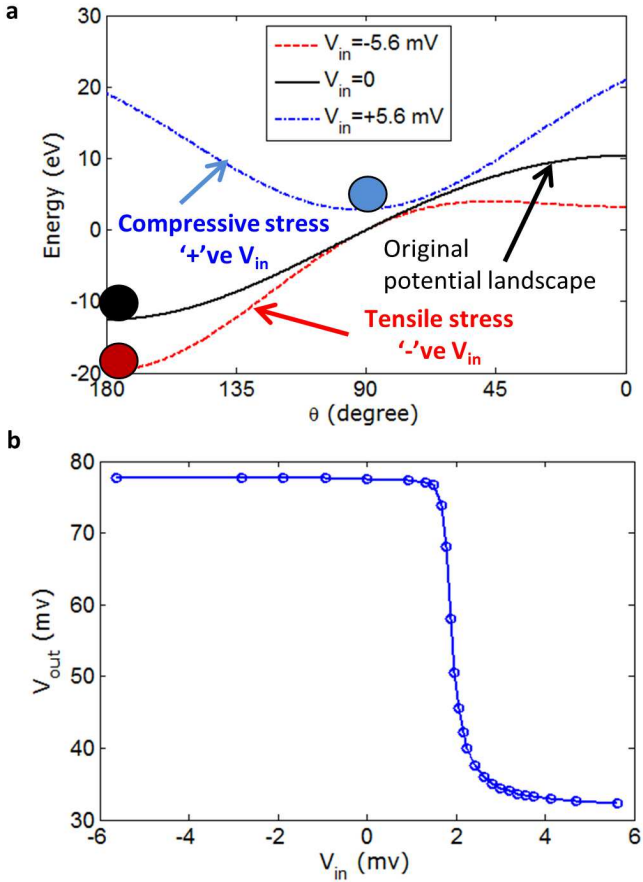}
\caption{\label{fig:roy_fig5} {A rectifier characteristics using strain-mediated multiferroic composites.} 
\textbf{(a)} Potential landscapes of the magnetostrictive nanomagnet when input voltages of different polarities are applied. 
\textbf{(b)} Stochastic LLG simulation results for the rectifier characteristics, i.e., input voltage versus output voltage characteristics when input voltages of different polarities are applied. Two different voltage levels are obtained depending on $V_{in} < 2$ mV and $V_{in} > 2$ mV.
}
\end{figure}

Figure~\ref{fig:roy_fig5} shows a characteristics of a rectifier using the strain-mediated multiferroic devices. Fig.~\ref{fig:roy_fig5}a depicts the operating principle of building the rectifier. The magnetization of the magnetostrictive nanomagnet gets more confined in the potential landscape when a negative input voltage is applied, while a sufficiently high positive input voltage makes the potential landscape monostable around $\theta=90^\circ$ as depicted earlier in Fig.~\ref{fig:roy_fig3}. Fig.~\ref{fig:roy_fig5}b shows the stochastic LLG simulation results of the rectifier depicting two different output voltage levels when input voltages of different polarities are applied. The degree of rectification increases with TMR. Such DC voltage level conversion may be required for different signal processing tasks~\cite{razav01,rabae03}.

\section{Conclusions}
\label{sec:conclusions}

In conclusion, we have demonstrated an intriguing feature of harnessing analog nature in inherently digital nanomagnets utilizing energy-efficient piezoelectric-magnetostrictive multiferroic composites. This eliminates the use of cumbersome and energy-consuming magnetic field, which can be applied along the hard-axis of a shape-anisotropic nanomagnet to rotate its magnetization. Different methodologies employing different ways to apply stress modulating the potential landscape of the magnetostrictive nanomagnets can be also possible to achieve the similar characteristics described here. Harnessing the transistor-like characteristics for analog signal processing tasks using these energy-efficient multiferroic devices apart from digital switching makes these devices very promising to be utilized in our future information processing systems.

\appendices
\section{Free layer's magnetization orientation versus output voltage relation}
\label{sec:theta_vout}

\begin{figure}
\centering
\includegraphics{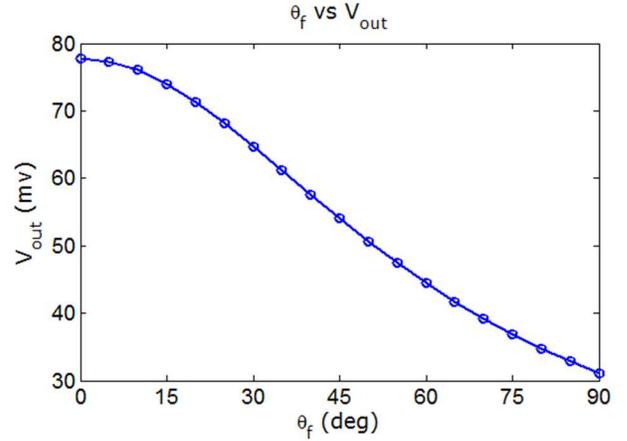}
\caption{\label{fig:results_MTJ} {The orientation of free layer's magnetization versus the output voltage from MTJ.} When free layer's orientation $\theta_f$ is close to $0^\circ$, the conductance is high since the free layer has antiparallel orientation with respect to the fixed layer. With the application of sufficient stress, the magnetostrictive nanomagnet (see Fig.~\ref{fig:potential_profile_dipole_coupled}) rotates by $\sim 90^\circ$ and the free layer's magnetization also rotates by $\sim 90^\circ$ since they are magnetically coupled; the conductance of the MTJ at such orientation is higher than that of the antiparallel orientation, and therefore the the output voltage $V_{out}$ is lower.}
\end{figure}

For the dipole-coupled structure as in the Fig.~\ref{fig:potential_profile_dipole_coupled} (Fig.~\ref{fig:roy_fig3}), the conductance of the magnetic tunnel junction (MTJ) can be written as a function of the free layer's magnetization orientation $\theta_f$ as~\cite{RefWorks:851}
\begin{equation}
G(\theta_f) = G_{90^\circ}(1-\eta\,cos\,\theta_f),
\label{eq:G_theta}
\end{equation}
where $G_{90^\circ} = G(90^\circ)$, $\eta=P_{free}P_{fixed}$, and $P_{free}$ ($P_{fixed}$) is the spin polarization of free (fixed) layer. The quantity $\eta$ can be expressed as 
\begin{equation}
\eta=\cfrac{G_P - G_{AP}}{G_P + G_{AP}},
\label{eq:eta}
\end{equation}
where $G_P=G(180^\circ)$ and $G_{AP}=G(0^\circ)$. The tunneling magnetoresistance TMR is expressed as 
\begin{equation}
TMR=\cfrac{G_P - G_{AP}}{G_{AP}},
\label{eq:TMR}
\end{equation}
and it is related to $\eta$ as 
\begin{equation}
\eta=\cfrac{TMR}{TMR+2}.
\label{eq:eta_TMR}
\end{equation}

Figure~\ref{fig:results_MTJ} shows the output voltage $V_{out}(\theta_f)=I_{read}/G(\theta_f)$ from the MTJ as a function of the free layer's magnetization orientation $\theta_f$ with the following parameters: TMR = 300\%, and resistance-area product $R_P A= 175\,k\Omega$-$\mu m^2$ (Ref.~\cite{RefWorks:33}), where $R_p=1/G_P$, and $A$ is the cross-sectional area of the junction. Although Ref.~\cite{RefWorks:33} uses CoFe and TMR=300\% is achieved at low temperature, using CoFeB the TMR can be as high as 604\% (Ref.~\cite{RefWorks:1346}). Instead of using CoFe/CoFeB or Fe for the fixed/free layers~\cite{RefWorks:33,RefWorks:74}, half-metals~\cite{RefWorks:581} can produce TMR as high as 1000\%. Also low resistance-area product as low as $R_P A= 4.3\,\Omega$-$\mu m^2$ (Ref.~\cite{RefWorks:786}) by lowering the MgO tunneling barrier thickness~\cite{RefWorks:74} is plausible. Note that we have used an elliptical cross-section for the cross-sectional area $A$ and we have utilized a read current $I_{read} = (\pi/4)1\simeq0.8$ nA. The voltage across the MTJ is only tens of millivolts and the bias dependence~\cite{RefWorks:851,RefWorks:15,RefWorks:14} of MTJ is not assumed. With bias dependence, TMR reduces a bit, however, with half-metals much higher TMR can be achieved anyway. Note that MTJ capacitance should be low enough so that it can follow high frequencies. There are experimental evidences of MTJ operation at more then 1 GHz~\cite{RefWorks:786,RefWorks:864} and spin-transfer-torque switching with 50 ps pulse using spin-valve and giant-magnetoresistance (GMR)~\cite{RefWorks:694}.

\section{Design of the fixed layer}
\label{sec:fixed_layer}

The fixed layer in the Figs.~\ref{fig:roy_fig2}a and~\ref{fig:roy_fig3}a is shown as a single layer for brevity, however, the design of such layer constitutes of multiple layers~\cite{RefWorks:300} and such structure is routinely experimentally realized to build magnetic tunnel junctions (MTJs)~\cite{RefWorks:786,RefWorks:773}. The fixed layer constitutes a synthetic antiferromagnetic (SAF) layer with a Ru spacer and a PtMn antiferromagnetic pinning layer~\cite{RefWorks:300,RefWorks:786,RefWorks:773}. A sample thickness values of such structure are PtMn(20 nm)/CoFe(2.3 nm)/Ru(0.85 nm)/CoFeB(2.7 nm) following Ref.~\cite{RefWorks:773}. Such structure can be experimentally probed to have a net zero (or a fixed amount of) effective field on the free layer (Fig.~\ref{fig:roy_fig3}).

\section{Input voltage versus stress anisotropy relation}
\label{sec:voltage_stress_anisotropy}

The stress anisotropy in a magnetostrictive nanomagnet (lying on $y$-$z$ plane, $x$ is the out-of-plane, see Fig.~\ref{fig:potential_profile_magnetostrictive}) can be expressed as 
\begin{equation}
E_{stress}= - (3/2) \, \lambda_s \sigma \Omega\,cos^2\theta,
\end{equation}
where $(3/2) \, \lambda_s$ is the magnetostrictive coefficient, $\sigma$ is the stress, $\Omega$ is the nanomagnet's volume, $\theta$ is the magnetization polar angle in standard spherical coordinate system.

For plane stress, the the out-of-plane $x$-component stress vanishes, making the in-plane ($y$-$z$ plane) strain components $\epsilon_{zz}=(\sigma_{zz}-\nu \sigma_{yy})/Y$, $\epsilon_{yy}=(\sigma_{yy}-\nu \sigma_{zz})/Y$, where $Y$ is the Young's modulus. Therefore,
\begin{equation}
\sigma = \sigma_{zz}-\sigma_{yy} = \cfrac{Y (\epsilon_{zz}-\epsilon_{yy})}{1+\nu}.
\end{equation}

Using $\epsilon_{zz}=d_{31} E$, $\epsilon_{yy}=d_{32} E$, and $E=V_{in}/t_{piezo}$ where $d_{31}$ and $d_{32}$ are the piezoelectric coefficients, $E$ is the electric field, $V_{in}$ is the input voltage, and $t_{piezo}$ is the piezoelectric thickness, we get the the relation between input voltage $V_{in}$ and stress $\sigma$ as
\begin{equation}
\sigma = \sigma_{zz}-\sigma_{yy} = \cfrac{Y (d_{31}-d_{32})}{t_{piezo}(1+\nu)} V_{in}.
\label{eq:voltage_stress}
\end{equation}
Therefore, with the material parameters defined in the Section~\ref{sec:mat_params_dimensions}, 1 mV of voltage can generate $\sim$5 MPa stress. Note that $d_{31}-d_{32}$ is negative, therefore a positive $V_{in}$ generates compressive (negative) stress.

We can also derive the relation between input voltage $V_{in}$ and stress anisotropy $E_{stress}$ as
\begin{equation}
E_{stress}= - (d_{31}-d_{32}) \cfrac{(3/2) \, \lambda_s Y}{1+\nu} \cfrac{1}{t_{piezo}} V_{in} \Omega\,cos^2\theta.
\label{eq:voltage_stress_anisotropy}
\end{equation}
Therefore, with the material parameters defined in the Section~\ref{sec:mat_params_dimensions}, 1 mV of voltage can generate $45$ kT (T=300 K) stress anisotropy. This is a huge anisotropy, since such value is similar to the potential energy barrier between two stable states of a nanomagnet and the error probability due to spontaneous switching of magnetization is as low as $e^{-45}$. 

\section{Energy dissipation}
\label{sec:energy_dissipation}

The energy dissipation consists of three components and due to: (1) Gilbert damping of magnetizations in the magnetostrictive nanomagnet and the free layer, (2) application of voltage on the piezoelectric layer, and (3) Ohmic dissipation in the MTJ stack. The equation to calculate the energy dissipation due to Gilbert damping has been already described in the Section~\ref{sec:stochastic_LLG}. Since magnetization does not switch, rather periodically rotates with the periodic input signal, such energy dissipation turns out to be 1.7e-4 aJ per cycle for 1 GHz sinusoidal signal.

Modeling the piezoelectric layer as a parallel plate capacitor, the capacitance C=1.25 fF and thus with the DC voltage $V_{dc}=1.8$ mV, the $CV_{dc}^2$ energy dissipation turns out to be 4e-3 aJ. This leads to miniscule energy dissipation in these multiferroic devices. For an AC input voltage with peak-to-peak voltage difference of $ V_{ac} = 18.75\,\mu$V (corresponding to 0.1 MPa stress), the $CV_{ac}^2$ energy dissipation is negligible.

The energy dissipation in the MTJ stack per cycle is $E_{MTJ}=\int_{0}^{t_{cycle}} I_{read}^2 R_{MTJ}(t) \, dt$, where $R_{MTJ}(t)$ is the resistance of the MTJ at time $t$. For $t_{cycle}=1$ ns, $I_{read} = 0.8$ nA, and MTJ parameters mentioned in the Section~\ref{sec:mat_params_dimensions}, the energy dissipation $E_{MTJ}$ turns out to be 0.033 aJ and the average power dissipation is 0.033 nW. This energy dissipation can be further reduced by designing the MTJ with a lower RA product. Therefore the total energy dissipation turns out to be less than 0.1 aJ per cycle.

\section{Additional simulation results}

\begin{figure*}
\centering
\includegraphics{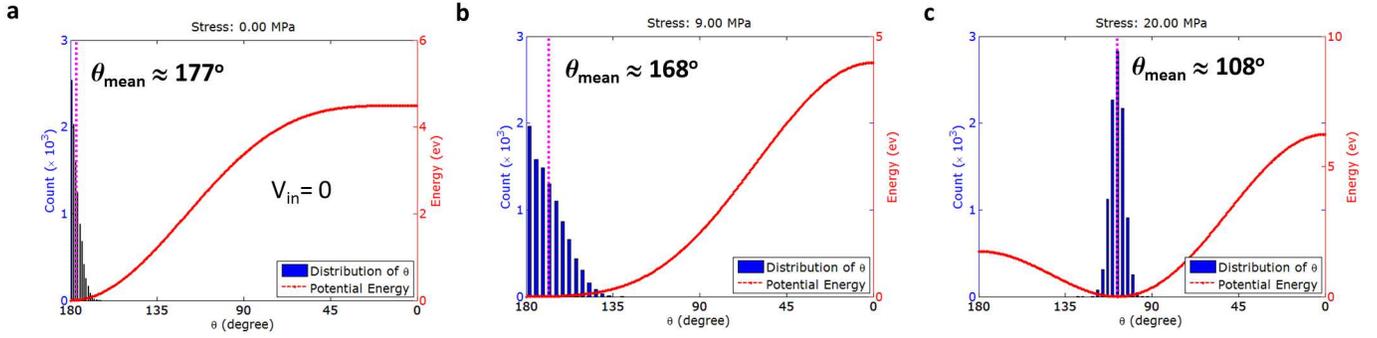}
\caption{\label{fig:results_magnetostrictive} {Continuous rotation of magnetization with stress.} 
\textbf{(a), (b), (c)} The results of stochastic LLG simulation for 0, 9 MPa, and 20 MPa stresses, respectively. Although the potential energy minimum does not change from $\theta=180^\circ$ for 9 MPa stress (case \textbf{b}), the distribution of magnetization becomes wider compared to the case \textbf{a} when no stress is active and the mean value of magnetization's orientation decreases. For case \textbf{c}, the stress is enough high to shift the magnetization's mean orientation close to $\theta=90^\circ$. At such sufficient stress and during the transition region, with a higher temperature, the standard deviation of the distribution increases, but the mean remains more-or-less constant. However, at lower stresses, when the barrier is not overcome by the stress yet, the mean deflection of magnetization increases with temperature, thereby decreasing the slope of the transition region. Hence, with increasing temperature, the gain tends to reduce.}
\end{figure*}

\begin{figure*}
\centering
\includegraphics{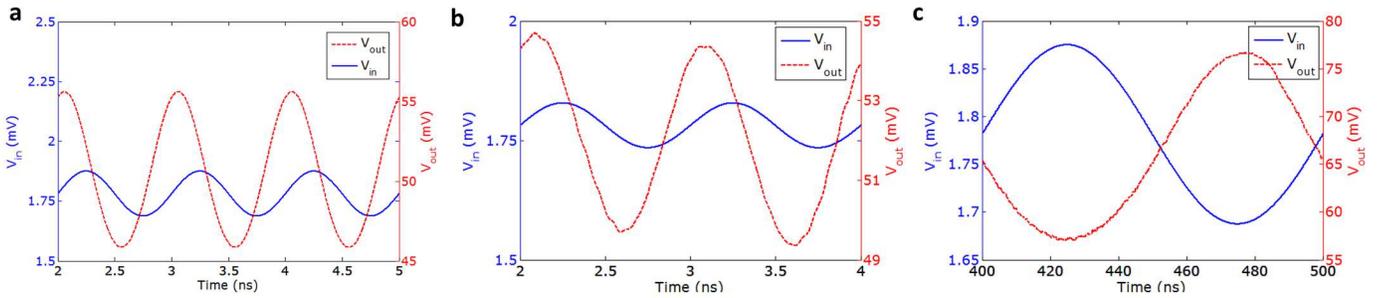}
\caption{\label{fig:results_dipole_coupled_input_output}
\textbf{(a)} An AC voltage with frequency 1 GHz is applied at the input terminal of the device in Fig.~\ref{fig:roy_fig3}a and the output voltage is calculated by solving Landau-Lifshitz-Gilbert (LLG) equation and then getting the output voltage from the TMR measurement. A voltage gain of 50 is achieved, while expending a miniscule amount of energy of less than 0.1 attojoule/cycle at room-temperature. As the input signal generates stress in the magnetostrictive nanomagnet and modifies the potential landscape of the nanomagnet, the effective field on the magnetization changes and follows the AC voltage. Therefore, the output voltage becomes a replica of the input voltage. Note that there is a phase difference between the input and output signals. No thermal noise is condsidered here.
\textbf{(b)} Input voltage frequency is 1 GHz, with thermal noise. Due to room-temperature (300 K) thermal fluctuations, there are occasional perturbations in the output voltage characteristics. 
\textbf{(c)} Input voltage frequency is 100 MHz, with thermal noise. Due to room-temperature (300 K) thermal fluctuations, there are occasional perturbations in the output voltage characteristics.The simulation time step is 1 ps, while the inverse of thermal attempt frequency $\Delta t=1$ fs. Note that there is a phase difference between the input and output signals. 
}
\end{figure*}

\begin{figure*}
\centering
\includegraphics{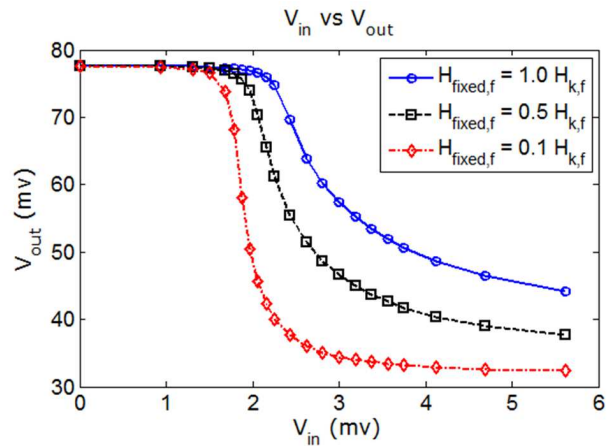}
\caption{\label{fig:results_dipole_coupled_Hz} {Variation of the slope of  $V_{in}$ versus $V_{out}$ characteristics with the effective field exerted by the fixed layer.} $V_{in}$ versus $V_{out}$ characteristics of the amplifier (Fig~\ref{fig:roy_fig3}) can be tuned by engineering the effective field exerted by the fixed layer. As the field is reduced, the slope of the high-gain region increases. Therefore, the maximum gain is achieved when no effective field is exerted by the fixed layer [synthetic antiferromagnetic (SAF) layer], i.e., net field acting on the free layer by the SAF layer is zero.}
\end{figure*}

Here, we provide some additional simulation results. The results are described from the corresponding captions of the Figs.~\ref{fig:results_magnetostrictive} --~\ref{fig:results_dipole_coupled_Hz} and they are referred from the main text.

\end{document}